\documentclass[a4paper,11pt]{article}
\usepackage{pos}
\usepackage{graphicx}

\newcommand{\pcal}{\mathcal{P}}

\title{Excited Hadron Channels in Hadronization}
\author*[a,b,c,d]{Rainer J.\ Fries}
\author[a,b]{Jacob Purcell}
\author[a,b]{Michael Kordell II}
\author[a,b]{Che-Ming Ko}

\affiliation[a]{Cyclotron Institute, Texas A\&M University,\\
  3366 TAMU, College Station TX 77843, USA}

\affiliation[b]{Department of Physics and Astronomy, Texas A\&M University,\\
  4242 TAMU, College Station TX 77843, USA }

\affiliation[c]{Frankfurt Institute for Advanced Studies,\\
    Ruth-Moufang-Strasse 1, 60438 Frankfurt am Main, Germany}

\affiliation[d]{ExtreMe Matter Institute EMMI,
   GSI Helmholtzzentrum für Schwerionenforschung, \\
   Planckstrasse 1,
   64291 Darmstadt,
   Germany }

\emailAdd{rjfries@comp.tamu.edu}
\abstract{
The proper treatment of hadronic resonances plays an important role in many aspects of heavy ion collisions. This is expected to be the case also for hadronization, due to the large degeneracies of excited states, and the abundant production of hadrons from their decays. We first show how a comprehensive treatment of excited meson states can be incorporated into quark recombination, and in extension, into Hybrid Hadronization. We then discuss the quantum mechanics of forming excited states, utilizing the Wigner distribution functions of angular momentum eigenstates of isotropic 3-D harmonic oscillators. We further describe how resonance decays can be handled, based on a set of minimal assumptions, by creating an extension of hadron decays in PYTHIA 8. Finally, we present first results by simulating $e^++e^-$
collisions using PYTHIA and Hybrid Hadronization with excited mesons up to orbital angular momentum $L=4$ and radial quantum number 2. We find that states up to $L=2$ are produced profusely by quark recombination.}

\FullConference{HardProbes2023\\
 26-31 March 2023\\
 Aschaffenburg, Germany\\}

\begin{document}
\maketitle

%\section{...}

Hadronization of partons is a longstanding problem for Monte Carlo (MC) event generators. The process
of forming bound states of quarks and gluons can not be described from first principles. Instead, several
models have been developed over the years to successfully describe certain aspects of hadronization. Among them
are the string fragmentation model, that is for example deployed in the PYTHIA event generator \cite{Andersson:1983ia,Bierlich:2022pfr}, and the quark
recombination model \cite{Fries:2003vb,Fries:2003kq,Greco:2003xt,Greco:2003mm}. String fragmentation is based on the idea of QCD strings 
forming between color charges at large distances, and has been successfully applied to all kind of "small" collision systems, i.e. those not involving 
nuclei. On the other hand, quark recombination has had success describing certain aspects of hadronization in nuclear collisions,
in particular large baryon/meson ratios and the constituent-quark-number scaling of elliptic flow. The idea behind Hybrid
Hadronization is to combine these two models so a comprehensive and consistent description of hadronization in all collision 
systems at a broad range of collision energies can be achieved \cite{Han:2016uhh,Fries:2019vws}. 

The overarching idea is that systems of partons that are ready to hadronize are first allowed to recombine by sampling
the recombination probabilities of all quark-antiquark pairs and all quark and anitquark triplets in the system. Gluons are assumed
to have decayed into quark-antiquark octet pairs for this step. The recombination
probabilities are computed in a phase-space formalism briefly sketched below. Some partons, preferentially those close in phase 
space, thus recombine directly into hadrons. All remnant partons are then assumed to be connected by strings and these string
are subjected to a string fragmentation model, in our case the one implemented in PYTHIA 8. If the original system of partons
has color tags assigned, e.g. in output from PYTHIA 8 for $e^++e^-$ or $p+p$ collisions, these color tags are used in the computation of recombination probabilites, and they are again used to form the remnant strings, preserving color flow in the parton system.

The focus of this work is the implementation of proper physical hadronic resonances in the recombination process. In the orignal work
\cite{Han:2016uhh} excited hadrons were not mapped onto the proper physical states in the Particle Data Book \cite{ParticleDataGroup:2020ssz} since the recombination probabilities into eigenstates of orbital angular momentum in the Wigner phase-space formalism were not known. Our study of 
angular momentum eigenstates of the 3-D isotropic harmonic oscillator \cite{Kordell:2021prk} remedies this shortcoming. We briefly discuss the work done
in \cite{Kordell:2021prk} and then present first results from an application to $e^++e^-$ collisions.

\begin{figure}[b]
\begin{center}
  \includegraphics[width=1.0\textwidth]{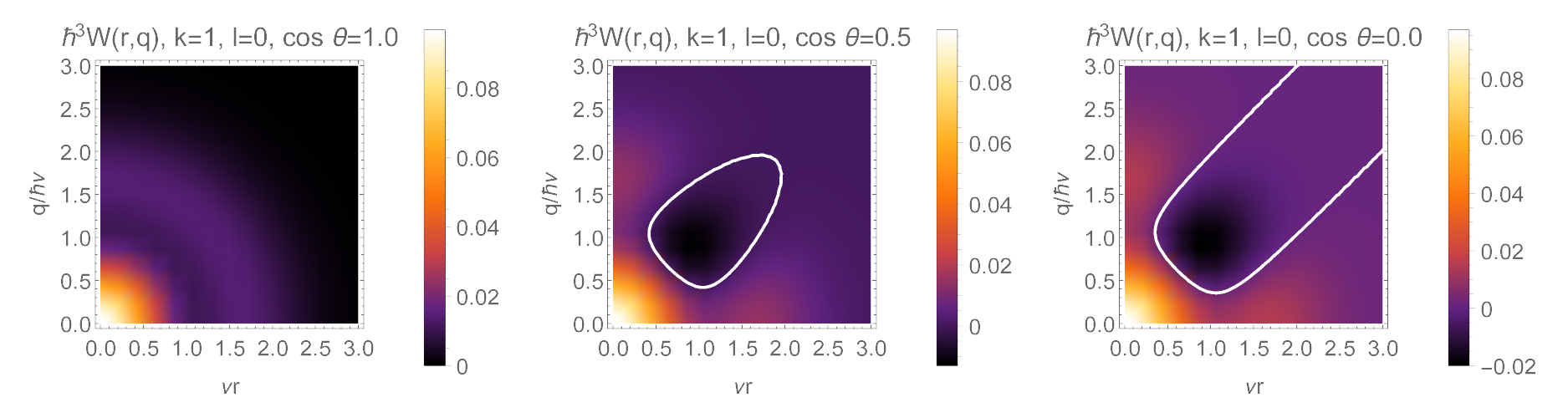}
\caption{\label{fig:wplots} Phase-space distributions $W_{10}$ for quantum numbers $k=1$, $l=0$ as functions of $r=|\mathbf{r}|$ and $q=|\mathbf{q}|$. The white lines indicate nodes where $W_{10}=0$. Distributions are shown for several values of the angle $\theta$ given by $\cos\theta = \mathbf{r}\cdot \mathbf{q}/rq$.}
\end{center}
\end{figure}

We assume the potential between color singlet quark-antiquark pairs to be be modelled by a 3-D isotropic harmonic oscillator. Our work restricts
itself to mesons for now for simplicity. The widths of the potentials can be fixed to data by computing the squared charge radii for stable mesons and
comparing those results to measured values of $\langle r^2\rangle$, as laid out in \cite{Han:2016uhh}. The inverse length scale of the harmonic oscillator is
denoted by $\nu$ in the following. We compute the Wigner distributions $W_{kl}(\mathbf{r},\mathbf{q})$ of eigenstates of the potential with 
radial quantum number $k$ and orbital angular momentum quantum number $l$. The magnetic quantum number $m$ is averaged over since we
do not wish to consider the polarization of hadrons. In that case 
the distributions only depend on the magnitudes of the position and momentum vectors $\mathbf{r}$ and $\mathbf{q}$, and the angle
$\theta$ between those vectors. In our work we reduce the complicated problem of 3-D phase-space distributions to the known Wigner 
distributions of the 1-D harmonic oscillator \cite{Curtright:2001}. These distributions have been computed before in \cite{Shlomo:1981ayz}
using a very different approach. The lowest energy distributions are
\begin{align}
  \label{eq:wigfinal}
  W_{00} & =  \frac{1}{\pi^3\hbar^3} e^{-\frac{q^2}{\hbar^2 \nu^2}-\nu^2 r^2 }\, ,  \\   
  W_{01} & =  W_{00}
     \left( -1 +\frac{2}{3} \nu^2 r^2 +\frac{2}{3} \frac{q^2}{\hbar^2 \nu^2} \right)\, ,  \\
  W_{10} & = W_{00}
     \left( 1 +  \frac{2}{3} \nu^4 r^4 -\frac{4}{3}\nu^2 r^2 %\right.  \\
    %& \qquad -   \left. 
   -\frac{4}{3} \frac{r^2 q^2}{\hbar^2} 
     + \frac{8}{3} \frac{(\mathbf{r} \cdot \mathbf{q})^2}{\hbar^2}        - \frac{4}{3} \frac{q^2}{\hbar^2 \nu^2} +  \frac{2}{3} \frac{q^4}{\hbar^4 \nu^4} \right)\, .
\end{align}
The distributions for $k=1$, $l=0$ are also visualized in Fig.\ \ref{fig:wplots}.

\begin{figure}[tb]
\begin{center}
  \includegraphics[width=\textwidth]{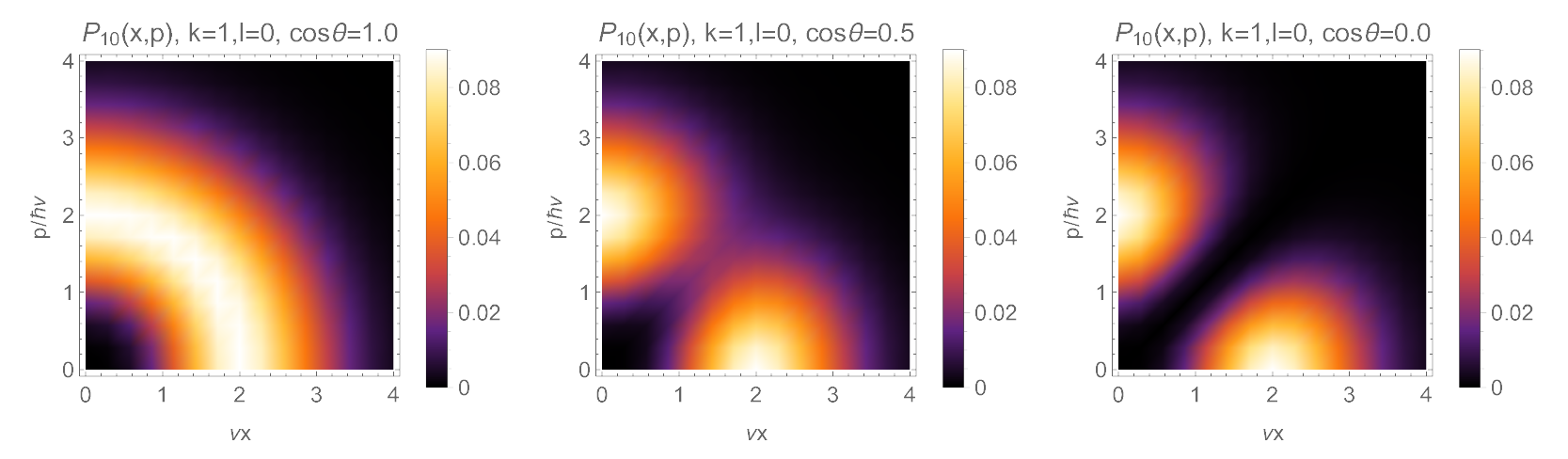}
  \caption{\label{fig:pplots} Coalescence probabilities $P_{10}$ for quantum numbers $k=1$, $l=0$ for two Gaussian wave packets interacting through an isotropic 3-D harmonic 
  oscillator potential. Probabilities are shown as functions of relative coordinates $x=|\mathbf{x}|$ and $p=|\mathbf{p}|$ for several values of the angle 
  $\cos\theta = \mathbf{x}\cdot \mathbf{p}/xp$ between the vectors. }
\end{center}
\end{figure}

We procede by computing the probabilites for two Gaussian wave packets with given widths $\delta$, representing the quark and antiquark, to 
coalesce into a bound state given by the Wigner distributions above. The probabilities
$P_{kl}(\mathbf{x},\mathbf{p})$ depend on the relative distance vectors $\mathbf{x}$ and $\mathbf{p}$ of the wave packets in position and momentum space, respectively. Here the magnetic quantum number $m$ is summed over to capture all polarization states. The probabilities for the 
lowest energy states are
\begin{align}
   \pcal_{00} &= e^{-u} \, ,  \\
   \pcal_{01} &= e^{-u} u \, , \\
   \pcal_{10} &= \frac{1}{2} e^{-u} \left( \frac{1}{3} u^2 -\frac{1}{3} t\right) \, 
\end{align}
for the simplest case $2\delta\nu=1$. For a discussion of other ratios between the length scales see \cite{Kordell:2021prk}.
The distrubutions have been written in terms of variables $u=\nu^2x^2/2+p^2/(2\hbar^2\nu^2)$ and $t=(\mathbf{x}\times\mathbf{p})^2/\hbar^2$, 
which have straightforward physical interpretations as the dimensionless squared distance of the initial partons in phase space, and their dimensionless 
squared angular momentum, respectively. It is then interesting to analyze the mapping of initial angular momentum $L^2$ in the parton system onto the choice of 
angular momentum quantum numbers $l$, in particular if states of several different $l$ with degenerate energy quantum number $n=2k+l$ are available \cite{Kordell:2021prk}. Maps of the coalescence probabilities $P_{10}$ are shown in Fig.\ \ref{fig:pplots}

We now turn to the implementation of this formalism for the recombination of mesons within the Hybrid Hadronization framework.
The phase-space probabilities $P_{kl}$ are supplemented with probablities $P_s$ to form spin single or triplet states and the probability 
$P_c$ to form color singlets. We have implemented light excited mesons up to $n=4$. This means we include mesons up to $G$-wave for 
the radial ground state $k=0$, and up to $D$-wave for a single radial excitation $k=1$. This set of states exceeds the states listed as confirmed 
by the Particle Data Group \cite{ParticleDataGroup:2020ssz}. To preserve unitarity we opt to keep the experimentally unconfirmed states. However, 
we are in need of the masses and strong decay branching ratios of these states. Missing masses are estimated by using the phenomenologically 
established scaling of squared masses with radial and spin quantum numbers, thus using linear interpolation and extrapolation. We consider decays 
into the simplest possible sets of stable hadrons plus additional pions, up to 
a maximum of five hadrons in total, which are allowed by the quantum numbers. Branching ratios are determined using phase-space weights and isospin algebra. As a random example, we pick the $1^3F_2$ isospin triplet state which would be known as $a_2(1918)$. We estimate its mass to be 
$1.918$ GeV and allow its decay into two 3-pion states (63.3\%) and three 5-pion states (36.7\%). The necessary data on the full set of meson states considered by us is collected in a XML particle data file following PYTHIA 8 standards, which enables Pythia to read in and decay the excited states
\cite{Kordell:2023xxx}.

With the formalism set we look at a first example. We use PYTHIA 8 to generate $e^+e^-$ di-jet events at $\sqrt{s}=91.2$ GeV. The partons after 
final-state showers are extracted and fed into a standalone version of the Hybrid Hadronization code that supports the generation of the new physical
meson resonances. Baryons are still treated as described in \cite{Han:2016uhh}. Hybrid Hadronization calls another instance of PYTHIA 8 to hand over remnant 
strings to fragment and includes hadrons from recombination. Both recombination and fragmentation hadrons can then decay according to settings in PYTHIA 8.

\begin{figure}[tb]
\begin{center}
  \includegraphics[width=0.3\textwidth]{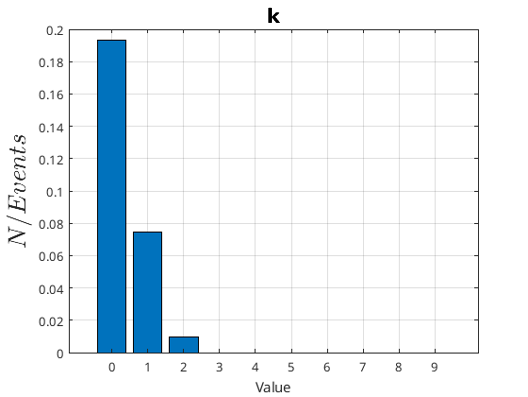}
  \includegraphics[width=0.3\textwidth]{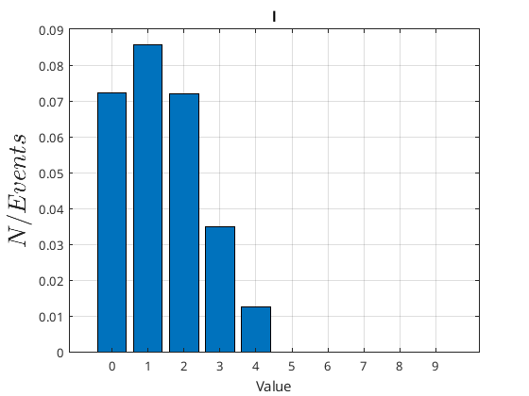}
  \includegraphics[width=0.3\textwidth]{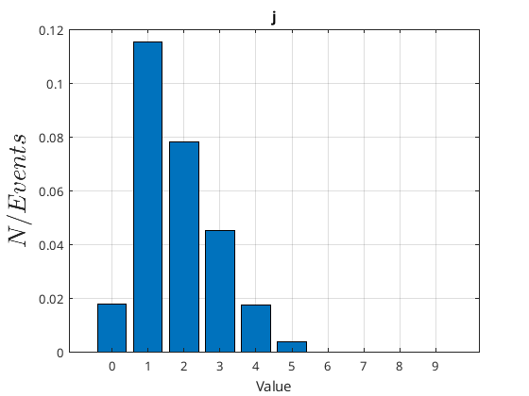}
  \caption{\label{fig:klj_dist} Relative yields of mesons of different radial quantum number $k$ (left panel), orbital angular momentum quantum number $l$ 
  (center panel) and total angular momentum quantum number $j$ (right panel) from recombination in $e^+e^-$ collisions at $\sqrt{s}=91.2$ GeV
  using PYTHIA 8 and Hybrid Hadronization, before decays of excited states.}
\end{center}
\end{figure}

We focus on one important preliminary result from this study. If one analyzes the quantum numbers of hadrons from recombination before any 
secondary decays. One finds the result shown in Fig. \ref{fig:klj_dist}. The three panels depict the relative abundances of mesons as functions
of radial and orbital angular momentum quantum numbers $k$ and $l$, as well as the total angular momentum quantum number $j$ which comes from the addition of
orbital and spin angular momentum operators, $\mathbf J =\mathbf{L} + \mathbf{S}$. The distribution of radial quantum numbers is strongly peaked 
at the ground state $k=0$ with the next excited stated adding about one third of that strength. On the other hand, $P$- and $D$-wave states 
($l=1$ and $2$, respectively) are as numerous as $S$-wave states, showing the importance of orbital angular momentum excitations in this jet systems. 
Even $F$-wave states are created with roughly half the strength of $S$-waves. As a result, vector and tensor mesons are, by far, the dominant channels in 
recombination, with $j=3$ mesons still being more than doubly as numerous as spinless mesons.

In summary, we have developed a formalism that allows for the inclusion of physical excited mesons into the quark recombination model and, 
in extension, the Hybrid Hadronization model. To this end we have computed the probabilties for the coalescence of Gaussian wave packets
into angular momentum eigenstates of 3-D isotropic oscillator potentials, approximating the dynamics of quark-antiquark systems at distances which
are not too large. We have utilized a phase-space formalism which allows us to compute quark-antiquark coalescence probabilities from the mean 
distance of the partons in coordinate and momentum space. We have compiled masses and branching ratios for strong decays for mesons up to 
$G$-waves, using estimates for those not found in the Particle Data Book. We find that in $e^+e^-$ collisions at $\sqrt{s}=91.2$ GeV, using PYTHIA 8
and Hybrid Hadronization, mesons states with excited orbital angular momenta dominate recombination.
In the future, we plan to extend our formalism to heavy flavor mesons and baryons.

This work was supported by the U.S.\ National Science Foundation under awards 1812431 and 2111568, 
and under award 2004571 through a subcontract with Wayne State University. In addition, this work was supported by the U.S.\ Department of Energy under Award No.\ DE-SC0015266. RJF would like to thank the ExtreMe Matter Institute EMMI at the GSI Helmholtzzentrum für Schwerionenforschung, Darmstadt, for support and
the Institute of Theoretical Physics at the University of Frankfurt and the Frankfurt Insitute for Advanced Studies for their hopitality.

\end{document}